\makeatletter
\def\cl@chapter{}
\makeatother
\documentclass[smallcondensed]{svjour3}   

\smartqed 
\usepackage{placeins}
\usepackage[utf8]{inputenc}
\usepackage[T1]{fontenc}
\usepackage{microtype}
\usepackage{balance}
\usepackage{fancybox}
\usepackage[english]{babel}
\usepackage{color}

\usepackage{url}
\usepackage{booktabs}
\usepackage{multirow} 
\journalname{Empirical Software Engineering (2019)}

\usepackage{caption}
\usepackage{graphicx} 
\usepackage{amsmath} 
\usepackage{amssymb}  
\usepackage{cleveref}
\usepackage{enumitem}
\usepackage{numprint} 
\usepackage[numbers]{natbib}
\usepackage{color}
\usepackage{listings}
\usepackage{parcolumns}
\definecolor{dkgreen}{rgb}{0,0.6,0}
\definecolor{gray}{rgb}{0.5,0.5,0.5}
\definecolor{mauve}{rgb}{0.58,0,0.82}
\newcommand{\eg}{{\textit{e.g.,}}}
\newcommand{\ie}{{\textit{i.e.,}}}
\newcommand{\ea}{{et al.}}
\newcommand{\cf}{{\textit{cf.,}}}
\newcommand{\aka}{{\textit{a.k.a.,}}}

\newcommand{\parrafo}[1]{\indent\textit{\textbf{#1:}}}

\newcommand{\reviewed}[1]{\textcolor{black}{{#1}}}
\newcommand{\hypobox}[1]{\begin{center}%
	\noindent\thicklines\setlength{\fboxsep}{10pt}%
	\cornersize{0.2}\Ovalbox{\begin{minipage}{4.3in}%
	\textit{#1}\end{minipage}} \end{center}}	

\newcommand{\RQone}{Can developers tell the difference between automated and manual refactorings?
} 
\newcommand{\RQtwo}{Is there any difference between the perceived quality of automated and manual refactoring?
} 
\newcommand{\RQthree}{Do automated refactorings affect code understandability?
}

\newlist{myenumi}{description}{10}
\setlist[myenumi]{labelindent=\parindent, leftmargin=*, label=(\roman*), align=left}
\setlist[myenumi]{leftmargin=0pt}

\begin{document}

\title{RePOR: Mimicking humans on refactoring tasks. Are we there yet?}

\author{Rodrigo Morales$^1$ \and Foutse Khomh$^2$ \and Giuliano Antoniol$^2$}

\authorrunning{Empir Software Eng (2019)} 

\institute{Rodrigo Morales \\
              rodrigomorales2@acm.org
\\\\
Foutse Khomh \\
              foutse.khomh@polymtl.ca
              \\\\
              Giuliano Antoniol \\
              giulio.antoniol@polymtl.ca
\\\\
$^1$ Department of Computer Science and Software Engineering, Concordia University, Montréal, Canada\\
$^2$ Département de génie informatique et génie logiciel, École Polytechnique de Montréal, Montréal, Canada
}
\date{Received: April 2019 / Accepted: --}
\maketitle

\begin{abstract}
Refactoring is a maintenance activity that aims to improve design quality while preserving the behavior of a system.  Several (semi)automated approaches have been proposed to support developers in this maintenance activity, based on the correction of anti-patterns, which are ``poor'' solutions to recurring design problems. However, little quantitative evidence exists about the impact of automatically refactored code on program comprehension, and in which context automated refactoring can be as effective as manual refactoring. Leveraging RePOR, an automated refactoring approach based on partial order reduction techniques, we performed an empirical study to investigate whether automated refactoring code structure affects the understandability of systems during comprehension tasks.   (1) We surveyed 80 developers, asking them to identify from a set of 20 refactoring changes if they were generated by developers or by a tool, and to rate the refactoring changes  according to their design quality; (2) we asked 30 developers to complete code comprehension tasks on 10 systems that were refactored by either a freelancer or an automated refactoring tool. To make comparison fair, for a subset of refactoring actions that introduce new code entities, only synthetic identifiers were presented to practitioners. We measured developers' performance using the NASA task load index for their effort, the time that they spent performing the tasks, and their percentages of correct answers. Our findings, despite current technology limitations, show that it is reasonable to expect a refactoring tools to match developer code. Indeed, results show that for 3 out of the 5 anti-pattern types studied, developers could not recognize the origin of the refactoring (i.e., whether it was performed by a human or an automatic tool). We also observed that developers do not prefer human refactorings over automated refactorings, except when refactoring Blob classes; and that there is no statistically significant difference between the impact on code understandability of human refactorings and automated refactorings. We conclude that automated refactorings can be as effective as manual refactorings. However, for complex anti-patterns types like the Blob, the perceived quality achieved by developers is slightly higher.


\end{abstract}

\section{INTRODUCTION}

In 1950, Alan Turing developed a test to assess a machine's ability to display behavior equivalent to that of a human being~\cite{turing2009computing}. The evaluator (human) will be exposed to a blind conversation with a machine and another human, and by formulating questions (s)he will try to identify his interlocutor (i.e., whether it is the human or the machine). If the evaluator cannot distinguish between human and machine, the latter one is said to have passed the test. In this paper, we conduct an experiment inspired by the Turing test. We want to check whether RePOR our automatic software refactoring tool~\cite{RePOR_JSS_symred} can be as effective as human developers at least from the refactoring structure point of view. 
Indeed, one major limitation of RePOR, and existing refactoring tools, is the lacking of semantic and contextual information  of identifiers. To have a fair comparison in our study, for a subset of refactoring actions, we used synthetic identifiers also for human refactored code and asked developers to judge the soundness of refactoring actions and code structure. It is important to underline, RePOR evaluation goal was not to verify if the refactoring results were deemed useful or necessary by the original developers or if original developers would have performed the same code changes. The overarching goal was the verify if RePOR was capable to produce refactoring actions and code comparable to human changes quality wise.

To this aim (1) we presented practitioners with refactorings performed manually by developers and refactorings performed by RePOR, and asked them to identify the origin of the refactorings. We also asked them to judge the design quality of the refactorings. Next, (2) we asked practitioners to perform a series of comprehension tasks on (both manually and automatically) refactored code and we assessed their performance.


\parrafo{Context} Software systems age as the result of deviations of their original design due to the implementation of new features~\cite{Parnas94-ICSE-SoftwareAging}, changes in the business logic, etc. One way to combat software system deterioration is to perform maintenance activities continuously; correcting 
poor design choices (\aka~anti-patterns)~\cite{Brown98-AntiPatterns} for example through refactorings~\cite{fowler1999refactoring}. Refactoring is a software maintenance activity that aims to reorganize code structure without altering system's behavior~\cite{fowler1999refactoring}. In fact, agile methodologies like \emph{eXtreme Programming (XP)} encourages developers to interleave refactoring with their code tasks to ease software evolution. However, refactoring is a time-consuming activity because (1) developers have to identify software components that contain anti-patterns; (2) select the adequate refactorings to clean-up the code; and (3) select the best order to apply the refactorings determined in the previous step. Yet, the benefits of code refactoring to combat software aging can only be observed in the long term.  

Previous works have studied refactoring practice in academic and industrial settings, and found that refactoring is a common practice, that the refactorings manually performed by developers differ from those applied using tool support, and that tool support is underused due to a lack of awareness~\cite{griswold1993automated,murphyreftools2008,vakilian2012}. In the last decade, tools and frameworks supporting  refactorings  have been proposed~\cite{moghadamAutoRefDesig2012,Ouni201518,morales2016Tarf,Morales2016Recon}. However, to the best of our knowledge, little effort has been taken to evaluate the impact on code understandability of automated-refactoring compared to manual refactoring. As a consequence, many developers express natural reluctance about \reviewed{applying automated refactoring to their code bases}~\cite{le2012systematic}. To accept or reject the notion that automated-refactoring can replace manual one, we propose to submit the refactorings applied by human and machine to the scrutiny of software developers, who will act as judges, \reviewed{similar to a Turing test}, to assess the quality, pertinence, and understandability of automatically generated code structure. Our goal is to verify if RePOR  challenges the idea that automated refactorings can't compete with refactorings performed by humans, in terms of quality, according to human judges. 
  
\parrafo{Premise}
Refactoring is conjectured in the literature to improve the design quality of systems. Despite the large number of studies on refactoring summarized in Section~\ref{sec:relatedwork}, few studies have empirically investigated the impact of automated refactoring tools on program comprehension. Yet, program comprehension is crucial to practitioners responsible of performing software maintenance. Hence, a better understanding of the conditions in which automated- is as suitable as manual-refactoring can promote the development of more human-like tools that perform automation efficiently, reducing maintenance costs.\\
\parrafo{Goal}
We want to gather quantitative and qualitative evidences that automated-refactoring can succeed in improving the design quality of a system at least as well as a human being would do in similar conditions. \\
\parrafo{Study}
We perform two experiments: (1) a refactoring survey ($E_1$) where we invited developers from Java mailing lists, and technical groups on social networks, to differentiate refactoring code changes (that aimed to remove anti-pattern's instances) generated by a software tool (machine) from those generated by freelance developers (human). For simplicity, we called this test in the consecutive "Turing test", acknowledging the  differences on implementation, as a Turing tests typically require human interaction between human judges and machine in the form of natural language. The freelancers that refactored the code were hired from two well-known crowdsourcing marketplace websites\footnote{Freelancer.com and Guru.com}.  We also asked the participants of the refactoring survey to rank the refactorings according to their quality. (2) We conducted a series of code comprehension experiments ($E_2$) where we studied whether the refactoring code changes generated by tool are more difficult to understand than those generated by developers. In $E_1$, we surveyed 80 developers using code from 10 different Java systems \reviewed{(two study groups, manual and automated code changes)}. In $E_2$, we hired 30 more additional developers,  from the two aforementioned crowdsourcing marketplaces, to perform two comprehension tasks covering three out of the four categories of comprehension questions identified by~Sillito~\ea~\cite{sillito2008asking}. We measured the subjects' performance using: (1) the NASA task load index for their effort; (2) the time that they spent performing the tasks; and, (3) their percentages of correct answers.\\
\parrafo{Results}
From the collected data, we observe that: (1) the ability of developers to recognize automatically-generated refactoring changes depends on the types of anti-patterns removed. For example,  automatically generated refactoring changes to remove Blob, Spaghetti Code and Lazy Class anti-patterns are hard to distinguish, while those generated for correcting Long Parameter List and Speculative Generality anti-patterns are not. (2) Developers do not have preference between refactoring changes generated by humans or  by machine (i.e., automated tools), except for those that correct Blob classes. (3) There is no statistically significant difference between the impact of automated- and manual- refactorings on program comprehension.  


\parrafo{Relevance}
Understanding the context in which automated-refactoring is beneficial to developers and the cases where a human supervision is required is crucial from the points of view of both researchers and practitioners. For researchers, our results debunk the myth that automated-refactoring is not reliable and-or cannot be safely performed without human intervention. At the same time, they also highlight the need to develop approaches able to create identifiers  that are semantically and contextual dependent. For practitioners, our results build confidence in the adoption of automated-refactoring tools. We hope that our results serve as inspiration to both groups, and help them develop new tools and approaches to support software maintenance and evolution.\\  
\parrafo{Organization}
Section~\ref{sec:relatedwork} relates our study with previous works. Section~\ref{sec:experimentalDesign} describes the design of our empirical study. Section~\ref{sec:results} presents the study results, while Section~\ref{sec:discussion} discuss further the obtained results and their implications. Section~\ref{sec:threats} discusses threats to validity.  Finally, Section~\ref{sec:conclusions} concludes the paper and discusses avenues for future work.
\section{Related Work}
\label{sec:relatedwork}

Fowler~\cite{fowler1999refactoring} popularized the term refactoring, by defining some heuristics to improve the design of existing code while preserving its functionality.  Brown~\cite{Brown98-AntiPatterns} introduced the notion of anti-patterns as poor-design choices that hinders code evolution.  Opdyke~\cite{opdyke1992refactoring} is the first to formulate a set of pre- and post-conditions to automatize the refactoring of object-oriented systems. Mens~\ea~\cite{mens2004survey} published an extensive overview of existing research work of software refactoring, and there are still new works published every year. 

In this section, we summarize some of the works related to refactoring and its impact on design quality and code comprehension, and relevant studies that compared manual and automatic refactoring.

\subsection{Impact of Refactoring anti-patterns on design quality and code comprehension}
Deligiannis~\ea~\cite{DELIGIANNIS2004129} proposed the first quantitative study of the impact of anti-patterns on software development and refactoring.  Through a controlled experiment involving 20 students and two systems, they found that Blob classes hinder the evolution of software design and the subject's use of inheritance. This work did not evaluate the understandability of the code, neither the subjects' ability to successfully perform comprehension tasks on the systems studied.

Stroulia and Kapoor~\cite{xing2006refactoringPractice} investigated the impact of refactoring on metrics associated to size and coupling and found that those metrics improved after refactoring.

In another academic setting, Du Bois~\ea~\cite{Bois2006DoesGC} found that the refactoring of God Classes into a number of collaborating classes can improve code understandability. The participants were asked to perform simple refactoring to decompose God classes. They found that participants exhibited less difficulties to understand the refactored code.

Murphy~\ea~\cite{murphy2012HowWeRefactoring} performed an empirical study on refactoring practice and the use of refactoring tool support. By analyzing the commit history of more than $39,000$ Eclipse developers, with their interaction traces (using Mylyn plug-in), they found that: (1) refactoring tooling support is rarely used; (2) that 40\% of refactorings assisted by tools occur on batch; (3) developers prefer interleaving refactoring with other code activities; (4) the kind of refactorings performed with tools differs from the kind of refactorings performed manually. However, this work did not study the impact on code comprehension of refactorings performed by machine and human, and what is the take of developers on automatically refactored code. These questions that were left unanswered served as motivation to perform our study.

Abbes~\ea~\cite{5741260} performed an empirical study to determine the impact of Blob and Spaghetti Code anti-patterns on program comprehension. They performed three experiments, each of them with 24 subjects and three different Java systems. They measured the subjects performance using three metrics: NASA TLX, tasks' completion times, and percentage of correct answers. They found that the occurrence of one single instance of anti-pattern does not impact significantly the comprehensibility of the code, while the combination of two anti-patterns does impact comprehension negatively.

Moser~\ea~\cite{moser2008case} analyzed the impact of refactoring on productivity in agile environments. They measured developer's productivity by dividing lines of code and effort (measured in time). They reported a statistically significant increase in productivity after refactoring along with some improvements in complexity and coupling.

Kim~\ea~\cite{6802406} performed a case study in an industrial setting at Microsoft on the benefits of refactoring. They found  that developers perceive manual refactoring as expensive and sometimes risky, and that refactoring changes loosely match the literature definition of semantics-preserving code transformations. They also reported that the top 5 percent of preferentially refactored modules in Windows 7 reported higher reduction in the number of inter-module dependencies and several complexity metrics, but as a consequence larger increase in module's size compared to the remaining 95\% of the refactored modules.

All these studies corroborate the idea that refactoring anti-patterns improve code design quality in several dimensions. However, all of them focused on manual refactoring.

\subsection{Manual compared to automatic refactoring studies}
Negara~\ea~\cite{10.1007/978-3-642-39038-8_23} performed an empirical study to compare semi-automatic (IDE tool-assisted) refactoring and manual refactoring operations of 23 participants from the academy and industry in a controlled experiment. They reported that developers performed 11\% more refactorings manually than using tool support; less experienced developers use less tool support than experienced ones; that developers perform large refactoring changes (\eg~extract method) with and without tool support, and that the size of a refactoring is not a decisive factor.  

Sz\H{o}ke~\ea~\cite{7332494} performed a case study with a R\&D company to study the effects of semi-automatic refactoring on code maintainability. They corroborate the statement by Murphy~\ea~\cite{murphy2012HowWeRefactoring} that the quality of machine-generated code might differ from manual refactoring when developers are prompted to provide input to the tool to decide between a list of refactorings, and when they blindly select the default machine propositions. They conclude that companies could achieve a considerable increment of code maintainability by only applying automatic refactorings.  Note that they did not compare an equal number of manual and automatic refactorings, but commanded developers to perform refactoring manually and then developed a tool to reproduce manual refactoring behavior followed by developers. The level of automation of their proposed tool is not disclosed.

Note that none of the studies in this category compared the understandability of manual and automatic refactoring on the same source code, and--or developers' perceived quality of the refactored code.

\subsection{Evaluation of Refactoring process}

Kataoka~\ea~\cite{kataoka2002quantitative} proposed an approach to measure the effect of refactorings on the maintainability of the code, using 
coupling metrics.  Although, they find that coupling metrics can be useful to asses the degree of maintainability improvement of a refactoring change, they also recognized that the type of refactorings that can be evaluated using coupling metrics is limited, 

In a recent preliminary work by Arima~\ea~\cite{ArimaERANaturalness}, a novel technique is proposed to assess the \emph{naturalness} of refactored code automatically, using probabilistic language models. 
\emph{Code naturalness} is a numerical value that measures how natural a given word sequence is for the model. As oracle they used a curated dataset of commits belonging to JUnit, where the refactorings were clearly identified by the authors of the commits. Their approach achieved 68\% of accuracy on 28 refactoring operations. This study does not evaluate the impact of refactoring, but tried to propose a new metric to quantitatively measure how well does a refactoring matches the text content of a software system. In the future, this technique could serve for further comparisons of automatically and manually refactored code in a quantitative way, and to improve existing automatic refactoring tools.

\section{Experimental Design}
\label{sec:experimentalDesign}
We perform two experiments to assess the comprehension of source code by developers in the presence of five types of object-oriented anti-patterns: Blob (BL), Lazy Class (LC), Long Parameter List (LP), Spaghetti Code (SC) and Speculative Generality (SG). We chose these anti-patterns because they are are representative of poor design choices .  In fact, Palomba et al.~\cite{palomba2014they} found in a case study with developers from both industry and academia, that Blob class and Spaghetti Code are highly recognized and considered high severity design problems; Speculative Generality was perceived as problem of medium severity; Long-parameter list and Lazy class were considered low severity problem.  The rational to study Long-parameter list is that this anti-pattern type is deemed to affect code readability, and Lazy class to bloat code design unnecessarily. Additionally, previous works have proposed approaches to detect the  anti-patterns studied in this work~\cite{moha2010decor,marinescuDetection}. Experiment 1 is about (1) recognizing if a refactoring code change applied to remove an instance of the aforementioned anti-pattern types was performed by a developer, or an automatic tool. We also asked participants to rank the refactoring code changes according to their perceived quality. 
In Experiment 2, \reviewed{using a different pool of participants, we present them with code refactored by either developers or an automated tool}, and ask them to complete some comprehension tasks on the refactored code entities. The aim is to assess the comprehensibility of code after refactoring. 
In each experiment, we selected two instances of each anti-pattern type studied: one \textit{easy} and one \textit{difficult}. The level of difficulty is decided based on several object-oriented metrics, based on each anti-pattern type.

\subsection{Research Questions}
Our research questions stem from our goal of understanding the impact of automated refactoring on developer's comprehension. We state them as follows.\\ \emph{RQ1: \RQone} \\
\emph{RQ2: \RQtwo} \\ \emph{RQ3: \RQthree}  

\subsection{Hypotheses}
For RQ1, we test the following null hypothesis when subjects review refactored code changes.
\begin{myenumi}
	\item [$H_{0identify:}$] Developers correctly differentiate the refactoring changes generated by an automated tool from the ones generated by another developer.
\end{myenumi}

For RQ2, to examine differences in the perceived quality of the refactoring changes, we test the following null hypothesis. 
\begin{myenumi}
	\item[$H_{0rank:}$] Developers prefer manual refactoring  over automated refactoring.
\end{myenumi}

RQ3 is related to subjects performing code comprehension tasks on refactored code. We study the performance of the subjects along three dimensions: time to execute the comprehension task, effort measured using NASA task load index (TLX)~\cite{sharek2011useable}, and percentage of correct answers.
We test $H_{0performance:}$ There is no difference between the performance of developers performing comprehension tasks on code refactored by  developers compared to automatically refactored code. 


\subsection{Objects}
\reviewed{The selection criteria of the anti-pattern instances refactored in this study is based on the detection results of our own tool for detecting and refactoring anti-patterns RePOR~\cite{RePOR_JSS_symred} applied to the SF110 corpus~\cite{TOSEM_evaluation}. SF110 is a representative sample of 110 Java projects from \emph{SourceForge} Website\footnote{https://sourceforge.net/}, which is a popular open source software repository, with more than 500,000 projects and with more than 33 million registered users.  The projects in the SF110 corpus are packed into a common build infrastructure, including developers tests suites and automated-generated test suites previously validated by the authors of the corpus. This allows us to validate, to some extent, that a regression is not introduced in the code after refactoring either by our approach or by developers, when generating the objects of our experiment, and to have a common framework to import and build the systems (SF110 uses Ant build system to automate the building process).}

\reviewed{Once we collected the results of the anti-patterns detection, the first author of this work and a Master's student in our research lab selected two instances, for each anti-pattern type studied, that were deemed representative examples of anti-patterns based on the works of Brown~\ea~and Fowler~\ea~\cite{Brown98-AntiPatterns, fowler1999refactoring}. In case of disagreement, we came to the second author of this work for solving differences, and followed a conservative approach of discarding instances where we could not reach consensus. 
We are also interested to know if different refactoring types affect the perception of developers or not.  However, study only one single instance of each type would be insufficient as one system could be intrinsically easier or more difficult to refactor.  Hence, we opted for selecting one  easy and one hard system for each anti-pattern type.  In total we study 10 Java systems (two for each anti-pattern type) from different domains and sizes.  
We select the anti-pattern examples from different systems for both treatments (automated and manual), to control for possible learning effect on the respondents, who might get familiar with the system's   code design. }
 We decide on the level of difficulty of the systems based on different metrics that reflect effort required to perform the refactoring.  In the case of the Blob, we measure the size of the Blob class (lines of code) and the number of \emph{data classes} associated to the Blob class; in this work, a \emph{data class} is a class composed mainly by attributes, and the only methods declared are accessors (\emph{getters} and \emph{setters}). With respect to Collapse Hierarchy and Lazy Class, we measure the \emph{number of incoming invocations (NII)}, as we suggest that removing these classes requires to update all method calls to the inlined classes, which in turn require extra effort when values of NII are large. For Spaghetti Code, we measured \emph{Mccabe Complexity (CC)}. Finally, for Long-parameter list classes, we consider the number of parameters.

The anti-patterns types studied are briefly introduced in~\Cref{table:defects}.  We provide the name, a brief description and the refactoring strategies used to remove them according to the literature of anti-patterns~\cite{fowler1999refactoring,Brown98-AntiPatterns}.
\begin{table}[t]
\centering
\renewcommand{\arraystretch}{1.0}
\renewcommand{\tabcolsep}{.5mm}
\caption{List of studied Anti-patterns types and the refactorings used to correct them.}
\label{table:defects}
\scriptsize
\begin{tabular}{|p{0.2\columnwidth}|p{0.43\columnwidth}|p{0.34\columnwidth}|}
\hline
Name &\ Description & Refactoring(s) strategy\tabularnewline
\hline
Blob (BL)~\cite{Brown98-AntiPatterns} & A large class that absorbs most of the functionality of the system with very low cohesion between its constituents. In addition, Blob Classes are surrounded by classes who serve mainly as data holders, and that does not implement any functionality (\aka~data classes).  & \textit{Move Method (MM)}. Move the methods that does not seem to fit in the Blob class abstraction to more appropriate classes~\cite{sengSB06}.  Another strategy, when there are not suitable classes to move methods,  is the creation of new classes with methods and attributes that have high cohesion, and that are semantically related (\aka~extract class refactoring).\tabularnewline
\hline
Lazy Class (LC)~\cite{fowler1999refactoring} & Small classes with low complexity that do not justify their existence in the system. & \textit{Inline Class (IC)}. Move the attributes and methods of the LC to another class in the system.\tabularnewline
\hline
Long Parameter List (LP)~\cite{fowler1999refactoring} & A class with one or more methods having a long list of parameters, specially when two or more methods are sharing a long list of parameters that are semantically connected. & \textit{Introduce Parameter Object (IPO)}. Extract a new class with the long list of parameters and replace the method signature by a reference to the new object created. Then access to this parameters through the parameter object. \tabularnewline
\hline
Spaghetti Code (SC)~\cite{Brown98-AntiPatterns} & A class without structure that declares long methods without parameters.
& \textit{Replace Method With method Object (RMWO)}. Extract long methods into new classes so that all local variables become attributes on that object.\tabularnewline
\hline
Speculative Generality (SG)~\cite{fowler1999refactoring} & There is an abstract class created to anticipate further features, but it is only extended by one class adding extra complexity to the design. & \textit{Collapse Hierarchy (CH)}. Move the attributes and methods of the child class to the parent and remove the \textit{abstract} modifier. \tabularnewline
\hline
\end{tabular}
\end{table}

\reviewed{In~\Cref{table:systems} we present information about the systems studied. The ID column contains the ID assigned by SF110 (we provided it as reference) to a system, system's name, description, use of synthetic identifiers for new code entities (S.N.), number of classes in the system (N.C.), anti-pattern type (Ap. Type), Lines of Code of source class containing the anti-pattern instance (LOC), level of difficulty according to the authors assessment (i.e., the Difficulty column), and the normalized static entropy (N.E.)~\cite{5070510} of the refactoring change for machine (N.E.M) and human patches (N.E.H).  S.N. column can take any of this  values: A (all refactoring examples), H (just human example), and  N (none  of the examples).  The normalized static entropy is computed using Equation 1.}

\begin{align}
	N.E.(code \ change) = -\sum_{k=1}^{n}\left(p_{k} {\ast} \log_{n}p_{k}\right),
\end{align}
where $p$ is the probability of changing a file $k$, with $p_k\geq 0, \forall k \in 1,2,\ldots,n$ and  $\sum_{k=1}^{n} p_{k}=1$. 

\reviewed{We measure $N.E.$ to quantify the effort required to refactor an anti-pattern based on the number of changes made by a developer (respectively a tool). From \Cref{table:systems} we observe that the refactoring effort as measured with $N.E.$ is very similar between machine and human treatments for most cases.}

\begin{table}[t]
\centering
\renewcommand{\arraystretch}{1.0}
\renewcommand{\tabcolsep}{.5mm}
\caption{List of systems from where we extract the anti-pattern's instances studied }
\scriptsize
\label{table:systems}
\begin{tabular}{|l|p{0.1\textwidth}|p{0.25\textwidth}|l|r|p{0.1\textwidth}|r|l|r|r|}
	\hline
	ID & System & Description &S.N. & N.C. & Ap. Type & LOC & Difficulty & N.E.M. & N.E.H.\\ \hline
	110 & FireBird & A relational database manager &H &158 & Blob & 27 & Easy & 0.94 & 0.94 \\ \hline
	83 & Xbus & It is a central Enterprise Application Integration (EAI) &H & 159 & Blob & 595 & Hard & 0.89 & 0.93 \\ \hline
	52 & Lagoon & A XML-based framework for web site maintenance & N &53 & Collapse hierarchy & 32 & Easy & 0.97 & 0.97 \\ \hline
	86 & Advanced T-Robots & Fighting robots Arena &N &177 & Collapse hierarchy & 25 & Hard & 0.92 & 0.92 \\ \hline
	47 & dvd-homevideo & Application to burn DVDs &N &2 & Lazy class & 24 & Hard & 0.95 & 0.94 \\ \hline
	101 & SAP NetWeaver & Server Adapter for Eclipse &N & 154 & Lazy class & 46 & Easy & 0.88 & 0.89 \\ \hline
	103 & Sweet Home 3D & An interior design application &A & 2 & Long-parameter List & 712 & Hard & 0.86 & 0.89 \\ \hline
	104 & Vuze & BitTorrent Client &A &1740 & Long-parameter List  &187 & Easy & 0.85 & 0.88 \\ \hline
	81 & JavAthena & Online role playing game & A &31 & Spaghetti Code & 808 & Easy & 0.44 & 0.99 \\ \hline
	70 & EchoDep & Digital preservation application &A &64 & Spaghetti Code & 645 & Hard & 0.59 & 0.62 \\ \hline
\end{tabular}
\end{table}
Our benchmark of refactoring changes is comprised of two sets: 10 \emph{high-level} refactorings applied automatically by our approach (\emph{M}); and 10 \emph{high-level} refactorings applied by real developers (\emph{D}).  A {high-level} refactoring is a refactoring composed of more than one {low-level} refactoring~\cite{opdyke1992refactoring}. For example: Collapse Hierarchy, which is a {high-level} refactoring, is composed of the following {low-level} refactorings: one or more pull-up method/attributes; delete class; remove abstract modifier; update class references.

 For set \emph{M}, we refactored  the selected anti-patterns' instances, using the Eclipse plug-in implementation of RePOR\footnote{\url{https://github.com/moar82/RefGen}}. For set \emph{D}, we hired five experienced Java developer freelancers from two well-known marketplace websites (Guru.com and Freelancers.com). We provide them with the definition of the studied anti-patterns and the Fowler's catalog of refactorings~\cite{fowler1999refactoring}, to allow them to select the most adequate refactorings, according to their experience. \reviewed{One can argue that it would be more natural to collect refactorings from the original developers of the studied systems, as they are familiar with the code.  However, it would be hard to first locate, and then convince the original developers from open-source systems, as most of contributors of open-source projects are volunteers, without any contractual relationship with the project. Instead, we assume an hypothetical scenario where a new developer is integrating to a project, and is assigned the task of refactoring the existing code.}
\reviewed{To ensure that the refactoring changes written by freelancers does not alter the original behavior of the code in question, we asked them to validate that the project's unit tests provided in the benchmark pass and, if necessary, that they perform any pertinent update the unit tests to be valid.  For example, if the refactoring change implies moving a method, it is necessary to update the unit test method to point to the new location of the refactored method.
We checked that unit tests pass after refactoring to have certain confidence that no code regression was introduced on the systems studied.} 
	
\reviewed{Each freelancer performed two refactorings, and they were excluded from experiment 1 and 2 to avoid introducing bias in our results.}
	
\subsection{Subjects}
\label{subsec:subjects}
We invited participants to participate in experiment $E_1$ 
 through Java developers' mailing lists, and through the social network of the authors of this paper. In total we received 87 answers from which we filtered out responses that were incomplete, and those where participants who did not provide justifications for their choices. We ended up with 80 responses for this study. All participants were volunteers and could withdraw at any time from the study, for any reason. From the 80 responses, 57 participants declared software development as their main occupation, 7 participants declared their main occupation to be research, 6 work as software architects, 4 scrum masters, 4 students and 2 fall in the \emph{Other} category. Among these 80 participants, 37 declared to work on both open-source and proprietary systems, 34 declared to work on proprietary systems and only 9 on open-source. Fifty percent (40 out of 80) of participants declared to have more than 5 years of experience as developers. Eighteen participants have more than 2 and up to 5 years of experience, 14 participants have between 1 to 2 years of experience, and 8 participants have less than one year of experience.

For $E_2$, we hired 30 more additional Java developers from aforementioned two marketplace websites. To select a candidate, we perform an informal interview with the candidates, and we set a minimum of experience of one year developing Java software for the industry. To control for possible confounding factors, we registered the number of years of experience of each participant as professional developers, and asked them to provide their Java's level of confidence (using a Likert scale from 1 to 5).

\subsection{Independent variable}
The independent variable for the two experiments is related to who refactored an anti-pattern instance, i.e., its origin, and it is a binary variable stating whether the refactoring was performed by an automatic tool or by a developer.

\subsection{Dependent Variables}
In $E_1$, the first dependent variable is a binary variable stating whether or not the participant correctly identified the origin of the refactoring change. The second dependent variable is a categorical variable capturing the perceived quality of the refactoring solution proposed, on a scale of 1 to 5, where 1 is for ``Poor'' quality, and 5 is for ``Outstanding'' quality. 


In $E_2$, the dependent variables measure the subjects' performance, in terms of effort, time spent, and percentage of correct answers. 
We measure a subject's performance using the NASA Task Load Index (TLX).  TLX evaluates the subjective workload of subjects.  It is a multidimensional measure that provides an overall workload index based on a weighted average of ratings on six sub-scales: mental demands, physical demands, temporal demands, own performance, effort and frustration.  We combine weights and ratings provided by the subjects into an overall weighted workload index by multiplying ratings and weights; the sum of the weighted ratings divided by fifteen (sum of the weights) represents the effort~\cite{sharek2011useable}. To measure the time that participants spent on each task, we recorded the participants' remote session on the virtual machine where participants performed their assignment. The time reported only considers the time spent on the comprehension task, from the moment they open the project in the IDE, until they close it. We compute the percentage of correct answers for each question by dividing the number of correct elements found by a participant by the total number of correct elements (s)he has found. For example, if the question requires to find the total number of code references for a given object, and  there are ten references but the subject finds only four, the percentage of correct answers is forty for that question.

\subsection{Questions}
For $E_1$, we use an online survey system (Jotform\footnote{http:www.jotform.com}) where we present the refactoring change, the online link to repository of the system that contained the anti-pattern's instance, and the type of anti-pattern. To summarize the refactoring changes applied to the studied systems, we generate for each anti-pattern's instance refactored, a patch file using the diff command from the control version system (Git). A patch file contains the description of the changes made using diff notation, which is a unified format that developers and control version systems can understand. The link to the repository of the systems studied provide respondents with a complete reference of the refactoring changes applied. We also provide a link to the repository containing the original source code; for example, if they want to study deeper the impact of the refactorings applied, they could clone the repository and apply the patch. In Listing~\ref{lst:refactoring-patch} we show a fragment of a refactoring change from our online survey.

\begin{lstlisting}[,language=Java, caption=Fragment of a refactoring change from the online survey.,label={lst:refactoring-patch},escapeinside={(*@}{@*)},basicstyle=\scriptsize,numbers=left,xleftmargin=15pt,breaklines=true]
diff --git a/src/main/java/org/character/data/TXTCharacter.java b/src/main/java/org/character/data/TXTCharacter.java
index df86a66..964b164 100644
--- a/src/main/java/org/character/data/TXTCharacter.java
+++ b/src/main/java/org/character/data/TXTCharacter.java
@@ -6,6 +6,7 @@ import java.io.FileReader;
 import java.io.IOException;
 
 import org.character.data.config.CharConfig;
+import org.javathena.core.data.Clazz007382383094620344;
 import org.javathena.core.data.Friend;
 import org.javathena.core.data.Hotkey;
 import org.javathena.core.data.IndexedFastMap;
@@ -14,7 +15,6 @@ import org.javathena.core.data.PersistenteData;
 import org.javathena.core.data.Point;
 import org.javathena.core.data.ROCharacter;
 import org.javathena.core.data.Skill;
-import org.javathena.core.data.ROCharacter.JOB;
 import org.javathena.core.utiles.Functions;
 
 public class TXTCharacter implements PersistenteData<IndexedFastMap<Integer, ROCharacter>>
@@ -139,7 +139,7 @@ public class TXTCharacter implements PersistenteData<IndexedFastMap<Integer, ROC
 		    currChar.setName (mainCharSL[2]);
 
 		    tmpSplit = mainCharSL[3].split(",");
-		    currChar.setClass_(JOB.parseFromValue (Short.parseShort (tmpSplit[0])));
+		    currChar.setClass_(Clazz007382383094620344.parseFromValue (Short.parseShort  (tmpSplit[0])));
 		    currChar.setBase_level  (Integer.parseInt (tmpSplit[0]));
 		    currChar.setJob_level  (Integer.parseInt (tmpSplit[0]));
\end{lstlisting}

In diff notation, a patch does not show the complete file, but only shows the code fragments that were modified.  Those code fragments are called chunk. The lines starting with "+" indicate new lines added, and "-" lines removed.  "@@" is the chunk header, where Git indicates which lines were affected.  For example, line 5 in Listing~\ref{lst:refactoring-patch} indicates that from file A  (original source code represented by a "-"), 6 lines are extracted starting from line 6.  From file B (refactored source code represented by a "+" ), 7 lines are displayed, starting also from line 6 .  The text after "@@" serves to clarify the context.  Git tries to display a method name or other contextual information of where this chunk was taken from in the file.


%

 We asked developers to answer whether the refactoring change was generated by a developer or a software tool. We also had an option ``Unknown'' that participants could select if they were unable to tell whether the refactoring change was performed by a developer or generated by a tool. To control for possible randomness in the answers, we asked participants to provide their level of confidence in their answers (on a scale of 1 to 5), and a brief explanation to support each answer. Since the quality of developers' solutions may be different from that of our automated approach, we asked participants to rate the quality of the refactoring changes (according to their perception of quality) and mark their preferred solutions. We also asked them to provide 
any additional comment about the refactoring change's quality, if they considered it appropriate.

For $E_2$, we used comprehension questions to elicit comprehension tasks and collect data on the subject's performance. As in~\cite{5741260}, we consider questions in three of the four categories of questions regularly asked and answered by developers~\cite{sillito2008asking}: (1) finding a focus point in some subset of the classes and interfaces of some source code, relevant to a comprehension task; (2) focusing on a particular class believed to be related to some task and on directly-related classes; (3) understanding a number of classes and their relations in some subset of the source code; and, (4) understanding the relations between different subsets of the source code. Each category contains several questions of the same type.

We only selected questions in the first three categories, since the last category concerns different subsets of the source code, while in our experiments, we focus exclusively on one or two packages at most, that are affected by a particular anti-pattern. For each category, we choose the two most relevant questions through discussions between the first author and a Master's student intern who collaborated on this work. The decisions were validated by the second author. Selecting two questions for each category of question, which provided us with two data points from each participant. The six selected questions are the followings.  The text in bold is a placeholder that we replace with the appropriate behaviors, concepts, elements, methods and types depending on the system on which the subjects performed their tasks.
\begin{itemize}
	\item Category 1: Finding focus points:
		\begin{itemize}
			\item Question 1: Where is the code involved in the implementation of \textbf{this behavior}?
			\item Question 2: Which type represents \textbf{this domain concept} or \textbf{this UI element or action}?
	  \end{itemize}
	\item Category 2: Expanding focus points:
	  \begin{itemize}
			\item Question 1: Where is \textbf{this method} called or \textbf{this type} referenced?
			\item Question 2: What data can we access from \textbf{this object}?
		\end{itemize}
		\item Category 3: Understanding a subset of classes:
	  \begin{itemize}
			\item Question 1: How are \textbf{these types or objects} related?
			\item Question 2: What is the behavior that \textbf{these types} provide together and how is it distributed over \textbf{these types}
		\end{itemize}
\end{itemize}
For example, with system \reviewed{\emph{83 Xbus} (\cf~\Cref{table:systems}), we replace ``\textbf{this behavior}'' in question 1, category 1, by ``manipulating and  store the email received by class \texttt{POP3XMLReceiver}''.}
For category 2, we acknowledge that the questions might be answered by developers using the IDE search functionality.  However, developers still must identify and understand the classes or methods that they consider relevant to the task.  Additionally, discovering classes and relationships that capture incoming connections prepare the developers for the questions of the third category.  \reviewed{Below we present the comprehension task for system \emph{83 Xbus}. } \\
\fbox{
\noindent\begin{minipage}[t]{\textwidth}
	Answer the following comprehension questions related to the system Xbus.
	We focus on the package \textbf{net.sf.xbus.technical.mail}.
	
	\begin{enumerate}
		\item Where is the code involved  in  manipulating and storing the email received by class \texttt{POP3XMLReceiver}?
	\item Which type (class) defines a method  for reading  an email after registering its receiver in the Transaction manager?
	\item 	Where is the type \texttt{POP3XMLReceiver} referenced?
	\item 	What data can we access from the object \texttt{mEmailMessage}?
	\item How are \texttt{Email} and \texttt{POP3XMLReceiver} related?
	\item What is the behavior that \texttt{POP3XMLReceiver}  and  \texttt{Email} provide together and how is it distributed over these types?
	\end{enumerate}
	
\end{minipage} }

\subsection{Anonymization of new code lexicon}
\label{subsec:anonymization}
Identifier names and comments (code lexicon), when thoughtfully assigned, can support developers to better understand a software system. However, as already stated, current existing automatic refactoring tools and frameworks  are not equipped with mechanisms allowing them to generate a human-like name for a new class, or a new method introduced when refactoring. On the other hand, developers generate names that reflect the roles of the entities and--or follow projects guidelines. To ensure a fair experiment using the refactoring changes generated by humans and by RePOR, we decided to post-process the changes by removing any code comment, and renaming any new class or method added to the code base with an artificial name (for both origins), to avoid providing any hint that can lead the human evaluators to discover the origin of the changes. Note that renaming new classes and methods was only necessary for the following refactoring types: Introduce parameter-object, Replace method with object, and Extract class, while Inline Class, Collapse Hierarchy, and Move method did not require it.

\subsection{Design}
For $E_1$, we divided the 20 refactoring patches (10 changes for each origin) into 4 groups. So each respondent will answer a survey containing 5 refactoring changes (each change corresponds to a different system) for removing 5 different types of anti-patterns. We also interleave the origin of the changes (machine, human) and the level of difficulty (easy and hard) in a way that any group has more than 3 patches from the same origin or level of difficulty. As previously stated, the level of difficulty is assessed using  well known object oriented metrics. To assess feasibility, that time given to respond the survey was enough, and anticipate adverse events, we performed a pilot study with two Post-doctoral fellows and one Ph.D. student from our lab, with more than 5 years of experience developing with Java.  The pilot study also helped us to refine the questions, and improve the visual design of our survey in terms of readability. Note that none of the people that participated in the pilot study took part to the final study.

In addition to asking respondents to identify the origin of the patch, we also asked them to justify their response in a free-text box, and to indicate how confident they felt when answering the questions, using a scale from 1 to 5, to control for possible randomness in the answers.

We present our design in~\Cref{table:e1design}. First column is the number of question and the rest of the columns corresponds to the different groups. Each cell contains the type of anti-pattern, using the abbreviations of Table~\ref{table:defects}, the ID of the system, and a letter indicating the origin of the patch (H:human, M: Machine).

\begin{table}[t]
\centering
\renewcommand{\arraystretch}{1.0}
\renewcommand{\tabcolsep}{.5mm}
\caption{$E_1$ Experimental Design}
\label{table:e1design}
\begin{tabular}{|l|l|l|l|l|}
\hline
Question & Group 1  & Group 2  & Group 3   & Group 4  \\ \hline
1     & SC-83-M  & BL-83-H  & BL-110\_M & LP-103-M \\ \hline
2     & CH-86-H  & SC-70-H  & SC-70-M   & SC-81-H  \\ \hline
3     & LC-101-M & CH-52-M  & LP-104-H  & LC-101-H \\ \hline
4     & BL-83-M  & LP-104-M & LC-47-0   & BL-110-H \\ \hline
5     & LP-103-H & LC-47-H  & CH-52-H   & CH-86-M  \\ \hline
\end{tabular}
\end{table}

For $E_2$, we use the same collection of refactoring examples from $E_1$, 5 anti-pattern types, and two instances for each type; each instance with two possibilities: being generated by human or by machine. That totals 10 refactoring changes for each origin.  
We hired 30 freelancers, who completed two tasks each of them, leading to 60 comprehension tasks.

\subsection{Procedure}
All the data collected is anonymous.  The subjects could drop the experiment at any time, for any reason and without penalty of any kind.  For the freelancers hired for refactoring the systems, and the second group hired for answering the comprehension tasks, we set milestones which clearly specify work deliverables, so we pay for each task completed.  All the freelancers hired for refactoring the anti-patterns studied passed an interview, where they stated their experience as Java developers, and refactoring, and correcting anti-patterns. In addition to the anti-patterns definitions and refactoring strategies, we provided them with references from the literature and web sites related to refactoring, but we did not persuade them to blindly follow any of these materials, but encouraged them to base their actions on their work experience and own reasoning.

For $E_1$, we first briefly introduced the description of the anti-patterns using~\Cref{table:defects}. Next, we asked them, to not base their judgment on code lexicon, and to accept that they will only focus on code structure and its quality, and not on indentations, naming conventions, space, etc.

For $E_2$, the subjects knew that they would perform comprehension tasks, but did not know the goal of the experiment nor whether the system was refactored by a developer or by an automated approach. We informed them of the goal of the study after they finished the experiment. 

\subsection{Analysis method}
In RQ1, to attempt rejecting $H_{0identify:}$, we test whether the proportion of refactoring changes correctly identified (or not) by participants, significantly varies between changes generated by human or by machine.  We use Fisher's exact test~\cite{sheskin2003handbook}, which checks whether a proportion vary between two samples.  We also compute the odds ratio ($OR$)~\cite{sheskin2003handbook} that indicates the likelihood for an event to occur. The odds ratio is defined as the ratio of the odds $p$ of an event occurring in one sample, \ie~the odds that changes generated by machine were correctly identified, to the odds $q$ of the same event occurring in the other sample, \ie~the odds that changes generated by humans were correctly identified: $OR=\frac{p/(1-p)}{q/(1-q)} $.  An odds ratio of 1 indicates that the event is equally likely in both samples. An $OR$ greater than 1 indicates that the event is more likely in the first sample (machine), while an $OR$ less than 1 indicates that it is more likely in the second sample (human).
In RQ2, we use a (non-parametric) Mann-Whitney test to compare the perceived quality (\ie{} the rates assigned by participants) of refactoring changes generated by machine with the rates of refactoring changes generated by humans. Non-parametric tests do not require any assumption on the underlying distributions. Other than testing the hypothesis, it is of practical interest to estimate the magnitude of the difference between the rates assigned to changes generated by machine and humans. Therefore, we compute the non-parametric effect size measure Cliff's $\delta$ ($ES$)~\cite{cliff2014ordinal}, which indicates the magnitude of the effect of the treatment on the dependent variable. The effect size is considered negligible if $< 0.147$, small if between $0.147$ and $0.33$, medium if between $0.33$ and $0.474$, and large if $> 0.474$~\cite{romano2006exploring}. 
For RQ3 and RQ4 we use Mann-Whitney test to compare two sets of dependent variables and assesses whether their difference is statistically significant. The two sets are the subject's data collected when they answer the comprehension questions on the  systems refactored by either machine or humans. For example, we compute the Mann-Whitney tests to compare the set of times measured for each subject on the systems refactored by either machine or humans. We also compute the Cliff's $\delta$ ($ES$).

\section{Study Results} 
\label{sec:results}
We now describe the collected data and present the results of our study, answering the four RQs formulated in~\Cref{sec:experimentalDesign}.

\subsection{RQ1: \RQone}

In~\Cref{table:rq1general} we summarize the results of $E_1$. In the first column we use the following abbreviations: \emph{all} for all the anti-pattern types studied, or the abbreviation of each type; columns 2 to 4 are the percentage of correct, wrong, or unknown answers; columns 5 to 7 are the corresponding percentages for machine changes (m), while 
columns 8 to 10 are the corresponding percentages for human changes (h).

When considering all anti-pattern types, we observe that the proportion of correct and wrong answers are the same (45\%). In the remaining 10\% of cases, respondents were not able to discern the origin of the changes, \ie{} \emph{I do not know} answer was selected, first row, columns 2-3. With respect to the anti-pattern's type  fixed, Long-parameter and Spaghetti code have the highest percentage of machine changes correctly identified (more than 70\%), and therefore we consider that they could not pass by manual changes. Conversely, Speculative generality, Blob and lazy class were correctly identified in less than 50\% of cases, indicating that to some extent they mimic human behavior on refactoring tasks.

If we study the percentage of correct answers by refactoring origin, we observe that respondents found it more difficult to identify human patches (38\%), whereas machine patches were little easier to identify (52\%). This trend holds for all anti-patterns studied, except for Speculative Generality (SG), where the percentage of correctly identified machine changes (23\%) is lower than the correctly identified human ones (55\%). This result is surprising since the refactoring type applied for removing the two different SG instances, which is Collapse Hierarchy (CH), was applied for both treatments. To find a plausible explanation for this result, we manually examined the cases where respondents failed to distinguish  refactoring changes corresponding to SG anti-pattern type, and observed that they belong to the easy instance, which corresponds to 14 different respondents who failed to identify the origin of the changes generated by the automatic tool to remove SG anti-pattern type. In general, human judges doubted the ability of software tools to perform this refactoring type. For example, one responded commented: \emph{``The abstract class is useless (the change is better). I don't think a tool can detect it.''}.

\begin{table*}[t]
\centering
\renewcommand{\arraystretch}{1.0}
\renewcommand{\tabcolsep}{.5mm}
\caption{$E_1$ RQ1 overall results}
\label{table:rq1general}
\begin{tabular}{|l|r|r|r|r|r|r|r|r|r|}
\hline
ap-type & correct & wrong   & unknown & correct-m & wrong-m & unknown-m & correct-h & wrong-h & unknown-h \\ \hline
all      & 45.00\%    & 45.00\%    & 10.00\%          & 52.00\%           & 38.50\%        & 9.50\%                    & 38.00\%         & 45.50\%      & 16.50\%                 \\ \hline
bl       & 30.00\% & 61.25\% & 8.75\%           & 37.50\%           & 57.50\%        & 5.00\%                    & 22.50\%         & 65.00\%      & 12.50\%                 \\ \hline
lc       & 47.50\% & 45.00\%    & 7.50\%           & 47.50\%           & 47.50\%        & 5.00\%                    & 35.00\%         & 42.50\%      & 22.50\%                 \\ \hline
lp       & 57.50\% & 33.75\% & 8.75\%           & 80.00\%           & 12.50\%        & 7.50\%                    & 35.00\%         & 55.00\%      & 10.00\%                 \\ \hline
sc       & 50.00\%    & 45.00\%    & 5.00\%           & 72.50\%           & 22.50\%        & 5.00\%                    & 42.50\%         & 37.50\%      & 20.00\%                 \\ \hline
sg       & 45.00\%    & 45.00\%    & 10.00\%          & 22.50\%           & 52.50\%        & 25.00\%                   & 55.00\%         & 27.50\%      & 17.50\%                 \\ \hline
\end{tabular}
\end{table*}

To control for some confounding factors, like development experience, or confidence when answering the questionnaire, we cluster the results based on developer's experience (Table~\ref{table:e1byexperience}) and developer's confidence for each patch reviewed (Table~\ref{table:e1byconfidence}).

We observed that the largest number of respondents declared to have more than five years of experience as Java developers, while the smallest group declared to have less than one year. Contrary to what one would expect, the group with more correct answers (51\%) was the group with one to two years of experience, followed closely by the group with more than five years of experience (48\%), indicating that the experience factor was not decisive to correctly identify the origin of the refactoring changes. The same situation occurs with the other groups, as the group with less than one year of experience correctly identified more refactoring changes than the group with more than two, up to five years of experience.  \reviewed{We suggest that perhaps developers surveyed with less experience, are more skilled in Java and more versed in object-oriented design, and beside their little experience, they have worked on more challenging projects than their more experienced peers, or they simply have more experience on code refactoring.}

Concerning the number of correctly identified refactoring changes by origin (columns 5-6, Table~\ref{table:e1byexperience}), we corroborate what we observed in the overall results, respondents had more difficulty to identify refactoring changes generated by human than by machine.  With one remarkable exception: developers with one up to two years of experience, who correctly identified more refactoring changes applied by human than by machine.

\begin{table}[t]
\centering
\renewcommand{\arraystretch}{1.0}
\renewcommand{\tabcolsep}{.5mm}
\caption{$E_1$ RQ1 results by respondent's expertise}
\label{table:e1byexperience}
\begin{tabular}{|l|r|r|r|r|r|}
\hline
Expertise (years)    & Total & Correct & Correct \% & Correct-h& Correct-m \\ \hline
\textgreater{}5      & 200   & 96      & 48.00\%    & 40            & 56              \\ \hline
\textgreater{}2 to 5 & 90    & 32      & 36.00\%    & 12             & 20              \\ \hline
1 to 2               & 70    & 36      & 51.00\%    & 20            & 16              \\ \hline
\textless{}1         & 40    & 16      & 40.00\%    & 4             & 12              \\ \hline
\end{tabular}
\end{table}

\begin{table}[t]
\centering
\renewcommand{\arraystretch}{1.0}
\renewcommand{\tabcolsep}{.5mm}
\caption{$E_1$ RQ1 results by respondent's confidence per question}
\label{table:e1byconfidence}
\begin{tabular}{|l|r|r|r|r|r|}
\hline
Confidence & Total & Correct & Correct \% & Correct-h & Correct-m \\ \hline
5          & 109    & 48      & 44.04\%    & 24        & 24        \\ \hline
4          & 142   & 69      & 48.59\%    & 29        & 40        \\ \hline
3          & 105    & 46      & 43.81\%    & 18        & 28        \\ \hline
2          & 30    & 12      & 40.00\%    & 3         & 9         \\ \hline
1          & 14    & 5      & 35.71\%    & 2         & 3         \\ \hline
\end{tabular}
\end{table}
With respect to confidence level declared per question, we observed that respondents felt confident enough, as in 60\% of the questions they selected a confidence level between 4 and 5. We also observed that  the largest proportion of correct answers (48.59\%) corresponds to the  confidence level of 4, followed by the ones identified with a confidence level of 5 (44.04\%); the next group are respondents with a confidence level of 3 (43.81\%); refactoring changes identified with a confidence level of 2, reached 40\% of correctly identified patches; the last group, the one with the lowest level of confidence, achieved the smallest proportion of correct answers (35.71\%), which makes sense. 

If we analyze results based on the origin of the refactoring changes (columns 5-6), we observe that the machine's ones are  more easily identified than the human ones, with one exception: those changes identified with a confidence level of 5 have the same proportion of correct answers. Contrary to the analysis based on respondents' expertise, the analysis based on respondents' level of confidence is more linear, 
as higher level of confidence led to higher number of correctly identified refactoring changes, in almost sequential order (the only exception being between levels 4 and 5, where respondents with confidence level of 4 achieved better results than those with confidence level of 5). \reviewed{Respondents that selected confidence level of 4 may have been more modest than those who selected 5. Still both intervals are on the top of the confidence scale.}

In Table~\ref{table:fisher} we present the contingency table for the number of correctly/incorrectly identified refactoring changes with respect to their origin: machine (m), human (h) and the results of the Fisher's exact test and odds ratio $OR$ between changes generated by human or machine. The contingency table shows the frequency distribution of the refactoring changes identified by the respondents with respect to the refactoring change's origin.  Fisher's exact test indicates whether a significant difference of proportions between  automatically- and manually-generated refactoring changes exists.  Odds ratio indicates the probability of a respondent to correctly identify a change according to the origin of the refactoring change analyzed.  

The first column of Table~\ref{table:fisher} corresponds to the anti-pattern's type; columns 3 to 4 correspond to the number of correctly/incorrectly identified changes by origin; and columns 5 to 6 reports the result of Fisher's exact test and $OR$s when testing $H_{0identify}$. In the first row, when aggregating all refactoring types together, the $p-value$ of the Fisher's exact test is statistically significant, with an $OR$ greater than one, indicating that changes generated by machine were easier to identify than the ones generated by humans.  Aggregating all anti-pattern types might not be very descriptive, specially considering that the refactoring types studied cover a wide range of design problems (coupling, lack of cohesion, design size, readability, etc).

To make a more precise analysis of the results, we expand our analysis to consider the refactorings of each anti-pattern type separately (rows 2 to 6, Table~\ref{table:fisher}).  We observe that only in two anti-pattern types, the results are statistically significant. Refactoring changes to remove long-parameter list have higher probability to be correctly identified by respondents when they are automatically generated (7 times more according to $OR$). Conversely, refactoring changes to correct Speculative Generality classes are more likely to be correctly identified when they are refactored manually ($OR<1$). Based on some respondent's comments, we suggest that the results obtained for Long-parameter list reflect respondent's view that the refactorings performed did not improve the quality of the systems. Because of this perception of poor quality, the respondents may have considered that such refactorings could have only been produced by an automatic tool. 
This results suggests that the human judges considered the introduce parameter-object applied by both,  developers and machine, artificial.  For example, Table~\ref{table:rq1general} shows that the largest proportion of correct answers for an anti-pattern type correspond to long-parameter list with 80\% of correct answers; conversely, 55\% of respondents incorrectly chose machine origin when the origin was human; the remaining 10\% of respondents  declared themselves unable to decide (they did not know). 
For Speculative Generality anti-pattern, respondents tend to attribute the refactoring changes proposed for removing this type to humans, and some of them even mentioned that the refactoring proposed was too complex to be generated by an automatic tool. Note that existing popular Java IDEs like Eclipse and IntelliJ IDEA provide refactoring support for Long-parameter list (introduce-parameter object refactoring) but not for Speculative Generality  (\eg~Collapse hierarchy refactoring). Hence, they could have been misled by the wide availability of automatic refactoring tool support for refactoring long-parameter list anti-pattern when making their choices. 

For the remaining anti-patterns types, the $p-values$ are $> 0.01$. Hence, we cannot reject $H_{0identify}$ for those types of anti-patterns. We therefore conclude that developers cannot differentiate between refactorings changes made by human and machine for those types of anti-patterns. 

\npdecimalsign{.}
\nprounddigits{2}
\begin{table}[t]
\centering
\renewcommand{\arraystretch}{1.0}
\renewcommand{\tabcolsep}{.5mm}
\caption{Contingency table and Fisher's exact test results for developers' survey on refactoring changes}
\label{table:fisher}
\begin{tabular}{|l|r|r|r|r|r|r|}
\hline
ap\_type & correct-m & wrong-m & correct-h & wrong-h & $p-value$ & $OR$ \\ \hline
all      & 104              & 96             & 76             & 124           & \textbf{<0.01}    & 1.77       \\ \hline
bl       & 15               & 25             & 9              & 31           & 0.22    & 2.05       \\ \hline
lc       & 19               & 21             & 14             & 26           & 0.36    & 1.67       \\ \hline
lp       & 32               & 8              & 14             & 26           & \textbf{<0.01}    & 7.22      \\ \hline
sc       & 29               & 11             & 17             & 23           & 0.01    & 3.51       \\ \hline
sg       & 9                & 31             & 22             & 18            & \textbf{<0.01}    & 0.24       \\ \hline
\end{tabular}
\end{table}

\hypobox{In general, automatically generated refactorings and refactoring changes made by humans were equally difficult to identify. In 10\% of cases, developers couldn't even make a decision and opted for the \textit{``I do not know''} option.  Based on anti-pattern types, the results vary depending on the type.   Refactorings generated for removing Long-parameter list and Speculative Generality anti-patterns failed the Turing test, whereas Blob, Lazy class and Spaghetti Code anti-patterns passed it.}



\subsection{RQ2: \RQtwo}
In this research question, we examine developers' perception of automated refactorings. We want to know whether they have the same appreciation for automatically generated refactorings and manual refactorings. The absence of statistically significant difference between the two treatments could serve as evidence that automated refactoring can be interleaved with manual changes without affecting the quality of a system.

\begin{table}[t]
	\centering
	\renewcommand{\arraystretch}{1.0}
	\renewcommand{\tabcolsep}{.5mm}
	\caption{$E_1$, RQ2 Median rates by anti-patterns' type}
	\label{table:user_rates}
	\begin{tabular}{|l|r|r|r|l|}
		\hline
		Antipattern Type & rate\_machine & rate\_human & $p-value$         & $ES$       \\ \hline
		all              & 3/5           & 3/5         & 0.02            & negligible \\ \hline
		bl               & 3/5           & 4/5         & \textbf{\textless{}0.01} & medium   \\ \hline
		lc               & 3/5           & 4/5         & 0.13            & small    \\ \hline
		lp               & 3/5           & 3/5         & 0.65            & negligible \\ \hline
		sc               & 3/5           & 3.5/5       & 0.10            & small    \\ \hline
		sg               & 4/5           & 3/5         & 0.38            & negligible \\ \hline
\end{tabular}\end{table}

\begin{figure*}[tp]
	\centering
	\captionsetup{justification=centering}
		\includegraphics[scale=0.45]{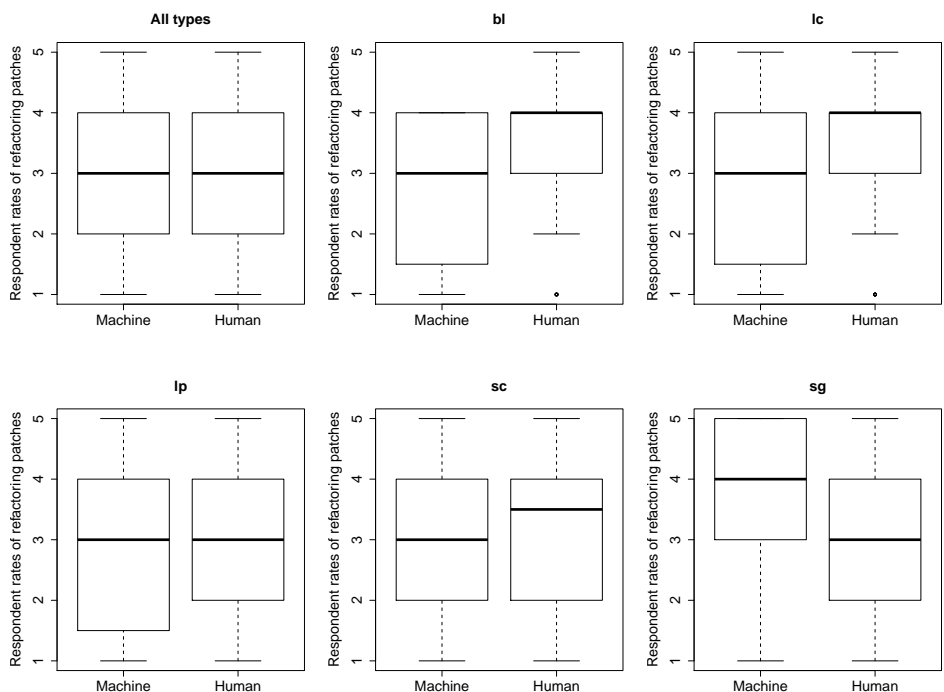}
	\caption{User's rates distribution. }
	\label{fig:user_rates}
\end{figure*}

In~\Cref{table:user_rates} we present the respondent’s median rates for the refactoring changes presented in $E_1$, \reviewed{and in~\Cref{fig:user_rates} the boxplots of user's rates distribution. }

Respondents input their rate for each refactoring change using a Likert-scale from 1 to 5. When considering all anti-patterns types, both sources of change (human, machine) attained the same rate. With respect to anti-pattern's type, the median rates for human changes are  higher (0.5 to 1 point) than the median rates of machine changes (Blob, Lazy class and Spaghetti code types). Conversely, the machine refactoring changes to remove Speculative Generality are rated higher than the changes generated by humans to remove the same anti-pattern type. Automatic and manual refactoring changes for removing Long-parameter list anti-patterns achieved the same rate.
We apply the Mann-Whitney test to determine if the results are statistically significant, and we find that only refactoring changes that remove Blob instances achieve statistical significance, with a $p-value$ less than 0.01, and a medium Cliff's $\delta \ ES$. Hence, we reject $H_0rank$ for all anti-patterns types except the Blob.

\hypobox{We conclude that for the anti-patterns types studied, developers do not find manual refactoring changes to be of better quality than automated refactoring changes. The exception being the Blob type.}

\subsection{RQ3: \RQthree}
\begin{figure*}[tp]
	\centering
	\captionsetup{justification=centering}
	\includegraphics[scale=0.45]{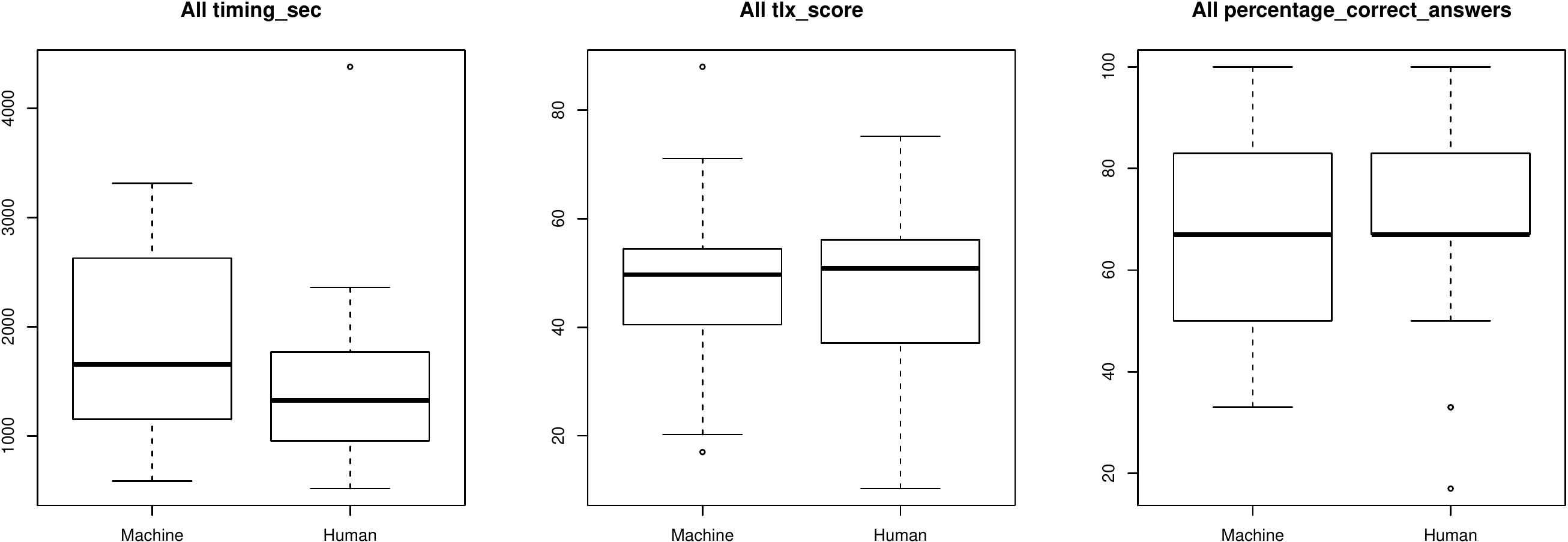}
	\caption{Comprehension tasks' performance results  distribution.}
	\label{fig:comprehension}
\end{figure*}

In this research question, we are interested to know the impact on understandability of refactoring changes generated by human and machine. 
We now present the results of $E_2$.  \Cref{table:comprehension} summarizes the median of collected data, \reviewed{and \Cref{fig:comprehension} presents the overall performance of participants on the comprehension tasks, in terms of: time, effort, and percentage of correct answers. In addition, we divide the refactoring changes in two groups (see rows 4 to 9 of \Cref{table:comprehension}), i.e., those who deceived human judges in $E_1$ (T. passed), and those who failed the test (T. failed).} In columns 5, and 6, we report the $p-value$ of the Mann-Whitney tests and the Cliff's $\delta \ ES$ obtained for each studied dimensions.

We observe that when considering all refactoring changes, the median time for completing the comprehension task on code automatically refactored is slightly higher than the median time spent on code manually refactored (approximately 5 minutes more). However, the difference is not statistically significant and the effect size is small. 

In the case of effort, the effort perceived by developers is almost the same, with a negligible effect size.

In the case of percentage of correct answers, the median values are the same for both human and machine.

We repeated the analysis along the same three dimensions while 
dividing the refactoring changes in two groups (passed and failed the Turing tests). With respect to time, the difference of the medians for anti-patterns that passed the Turing test (\ie~BL, SC, LC) is higher (medium $ES$) for automatically generated changes, though the difference is not statistically significant. In the case of effort, the effect size is negligible and not statistically significant, same for percentage of correct answers.  

For anti-patterns that failed the Turing test, the results are not statistically significant either.


\npdecimalsign{.}
\nprounddigits{2}
\begin{table}[t]
\centering
\renewcommand{\arraystretch}{1.0}
\renewcommand{\tabcolsep}{.5mm}
\caption{RQ3 Comprehension task results $E_2$: median values for the three dimensions studied}
\label{table:comprehension}
\begin{tabular}{|l|l|n{5}{2}|n{5}{2}|n{5}{2}|l|}
\hline
Grouping       & Dimension                    & Human  & {Machine} & {$p-value$}  & {$ES$}       \\ \hline
ALL            & Time (seconds)                  & 1328   & 1657.5  & 0.069845 & small    \\ \hline
ALL            & Effort (TLX index) & 50.895 & 49.705  & 0.822843 & negligible \\ \hline
ALL            & \% of correct answers & 67     & 67      & 0.760319 & negligible \\ \hline
T. passed & Time (seconds)                  & 1100   & 1606    & 0.087628 & medium   \\ \hline
T. passed & Effort (TLX index) & 51.365 & 49.705  & 0.720635 & negligible \\ \hline
T. passed & \% of correct answers & 67     & 75      & 0.829502 & negligible \\ \hline
T. failed & Times (seconds)                  & 1176   & 1606    & 0.159852 & small    \\ \hline
T. failed & Effort (TLX index) & 51.365 & 48      & 0.363437 & small    \\ \hline
T. failed & \% of correct answers & 83     & 67      & 0.011397 & large    \\ \hline
\end{tabular}
\end{table}

Since we could not find statistically significant differences in the performance of developers when performing code comprehension tasks on systems that were manually and automatically refactored, we reject $H_{0performance}$. Hence, we conclude that for the anti-patterns studied in this work there is no difference between the performance of developers performing comprehension tasks on code refactored by developers compared to automatically refactored code. This is good news for toolsmiths working on automatic tools for removing anti-patterns. For developers, this result will likely increase their trust in automatic refactoring tools, and for researchers it extends the existing body of knowledge on the benefits of automatic refactoring.

\hypobox{We conclude that for the set of refactoring strategies studied, developers can safely use automated tools since the impact of the automated changes on the comprehension of the code is not significantly different from the impact of changes performed by human developers.} 


\section{Discussion}
\label{sec:discussion}
In this section, we provide further information about the obtained results and discuss their implications. 
The results in $E_1$ show that in general, automatically generated refactorings and refactoring changes made by humans were equally difficult to identify. In 10\% of cases, developers couldn't even make a decision and opted for the \textit{``I do not know''} option. 

\subsection{\reviewed{Refactoring change differences by anti-pattern type}}
When analysing results based on anti-patterns types, the picture is a bit different. 
\subsubsection{Long-parameter-list anti-patterns automatically refactored were easier to identify than their manual counterpart}

For the two instances of Long-Parameter list studied, the refactoring strategy selected by both human and machine was to introduce a parameter-object. In fact, the solution provided for the \emph{easy} instance is conceptually the same for both the automatic tool and the developer (the only difference is the name selected for the new class). However, since we replaced the name in both solutions by an artificial name (\eg~Clazz09465) to avoid introducing bias in the experiment (as explained in Section~\ref{subsec:anonymization}), the two refactoring changes are semantically equivalent. In the \emph{hard} instance, there was a small difference between the two solutions; \reviewed{the freelancer} declared the new class in the same file, while our automated approach created a new file; besides that minor change, we can consider both solutions to be semantically equivalent too. 

As we mentioned in Section~\ref{sec:results}, we suggest that the quality of the refactoring changes applied to correct Long-parameter list did not influence the choice of the participants in the Turing test, but their expectations about refactoring. Some respondents mentioned that there is no reason to extract a new parameter object, given that the parameter object only stores data. Hence, we suggest that the developers surveyed did not give importance to the understandability aspect of having a long-list of parameters over having an extra class that clusters the common parameters together, and that is why they attributed the change to a machine. Note that the number of parameters extracted for the Long-parameter list instance identified by the authors as \emph{hard} is considerably long (8 parameters) compared to the rest of the classes (4 parameters) in the system where the instances of Long-parameter list anti-pattern resides. 

\subsubsection{Refactoring changes of speculative generality and lazy class are semantically equivalent}
With respect to Speculative Generality and Lazy class, the refactoring strategies employed by both the automated tool and the developers are the same for the first anti-pattern's type (\ie{} \textit{Collapse hierarchy}), while for the second type (\ie{} Lazy class) the refactoring applied (\ie{} \textit{inline class}) exhibits a small variation on the selected target class. The Lazy classes instances are composed of a unique static method, so they could be placed in any class that makes use of this method, so we consider the refactoring solutions for these two anti-pattern types to be equivalent, despite their origin. Given that in the solutions for both anti-pattern's types no classes or methods are added to the system, these refactorings can be completely automated, as they do not require to provide names to entities, according to some code conventions set for the systems.

\subsubsection{The hard instance of Spaghetti code that was manually refactored was easier to identify than the automatically refactored one}

\reviewed{The machine and human refactoring changes applied to remove spaghetti code, for the easy and hard instances, used the same strategy: \emph{replace long method with method object}. This strategy consists of extracting a long method inside the spaghetti code class into a new class.
For the easy instance, both solutions differ just in the name selected for the new class, and since we normalized the names, we  consider them semantically equivalent. On the other hand, the refactorings applied for the hard instance of spaghetti code class exhibit some design differences that may have help some survey respondents to identify their origin.}

\reviewed{The long method extracted from the spaghetti code class makes use of a private attribute and a static method inside the spaghetti code class, and it  does not require access to other external attributes or methods. In the automated refactoring change, two private attributes where moved to the new extracted class, while in the manual solution the freelancer added public getter to access the required private attribute outside the spaghetti code class. Hence, the use of the private attributes, in the automatically refactored code, are delegated to the extracted class inside the spaghetti code class. Since both machine and human set the access modifier of the static method to public, there was no need to pass a reference to the spaghetti code class in the automated refactoring solution. But in the manual solution the extracted long method requires a reference to the spaghetti code class to use the required private attribute. These differences were noticed by one of the survey respondent, who provided the following explanation to justify why he thought that the refactoring was automatically generated:\emph{``I don't think that a developer would extract the long method from the spaghetti code class to the new extracted class, and add as a parameter to the long method a reference to the spaghetti code class, because that reference is not used''}. With respect to the manual solution, one respondent commented that the addition of a getter, to the private attribute, and the way the extracted class was instantiated to call the long method does not seem to be automatically generated. To provide a complete picture of what we discussed, we show in Listing \ref{lst:refactoring-patch-sc-m} and Listing \ref{lst:refactoring-patch-sc-h}, the actual changes performed to instantiate the new class, and execute the extracted long method for both automatic and manual solutions. We use the acronyms ADD, DEL, and CHG to indicate added, deleted, and changed line.}
 
\noindent\begin{minipage}{.45\textwidth}
\begin{lstlisting}[language=Java, caption=Extract from refactoring change for \emph{hard} instance of spaghetti code by machine.,label={lst:refactoring-patch-sc-m},escapeinside={(*@}{@*)},basicstyle=\scriptsize,numbers=none,xleftmargin=15pt,breaklines=true,frame=tlrb]
public class WorkflowManager {
DEL	private String logconfig = System.getenv ("HS_HOME") + File.separator + "config" + File.separator + "log4j.properties";

ADD private Clazz008353 Clazz008353 = new Clazz008353 (System.getenv("HS_HOME") + File.separator + "config" + File.separator + "log4j.properties");

 	public WorkflowManager() {
DEL	PropertyConfigurator.configure (logconfig);
ADD	PropertyConfigurator.configure (Clazz008353.logconfig);
	}
	private void process_submission(String[] args) {
CHG	Clazz008353.process_submission (this, args);
 	}
 }
\end{lstlisting}
\end{minipage}\hfill
\begin{minipage}{.45\textwidth}
\begin{lstlisting}[language=Java, caption=Extract from refactoring change for \emph{hard} instance of spaghetti code by human.,label={lst:refactoring-patch-sc-h},escapeinside={(*@}{@*)},basicstyle=\scriptsize,numbers=none,xleftmargin=15pt,breaklines=true,frame=tlrb]
public class WorkflowManager {
     private void process_submission(String[] args) {
CHG new Clazz008353 (this).submit(args);
     }
 }
\end{lstlisting}
\end{minipage} 
 
 \reviewed{The differences between the automatic solution for the \emph{hard} instance of Spaghetti code with respect to the manual one could be easily matched by adding a validation to prevent passing a reference of the refactored spaghetti class as parameter to the long method call, when none of its attributes and methods are used. Toolsmiths should pay attention to corner cases like this, which can only be revealed by thoroughly testing their tools with several scenarios and systems.}

\subsection{Perceived quality of refactoring changes}

Concerning quality perceived by developers, we observed that it is only for the Blob anti-pattern type (which passed our Turing test) that the difference of rates achieved between manual and automatic refactoring changes is statistically significant, favoring manual refactoring.  

We \reviewed{identify} the following differences in the refactoring strategies \reviewed{used} by human and the automatic approach that are worth to discuss. 

First, the automated approach selected \emph{move method} as a mean to decompose Blob classes and to redistribute functionality to other classes in the system, while the human approach relied on extracting new classes to reduce the size of Blob classes. 

With respect to the refactoring strategy applied for the \emph{easy} instance of Blob,  the automated solution consists of delegating 12 methods to a class where an association relationship exists.  One respondent complained that the coupling between these two classes increased dramatically, which was not the case. \reviewed{In contrast,  the manual solution for the \emph{hard} Blob's instance took a different path}. It consists of extracting two attributes and their corresponding getters and setters to a new class.  This solution, though semantically correct, does not reduce the size of the Blob class significantly, but introduce coupling to a new data class. 

For the \emph{hard} instance, one respondent commented about the manual solution as follows: \emph{``it seems good and needs to be corrected with minor changes''}. We studied the manual solution and found that it performs a clear separation of concerns by extracting methods and attributes to new classes according to their names. That would be hard for a machine to achieve, unless they gather some knowledge about the code lexicon. The refactoring strategy followed by the automated approach consists of delegating three methods to a related class that is member attribute of the Blob class. \reviewed{  Although, the strategy taken by the tool prevents adding extra coupling to a new extracted class while reducing Blob's class size, respondents of $E_1$, did not like it. They highlighted that when moving methods, the automated approach changed the visibility of the required attributes to public, which implied several changes with no clear benefit.} 
\reviewed{Respondents rated the automatically-generated refactorings for Blob type} with average scores of 2.6 and 2.7, for the easy and hard instances respectively. While the human ones obtained 4.4 and 2.5 respectively. It is interesting to note that although the automatically-generated refactorings of Blob did not achieved the best rate, the perceived quality showed little variation between the two instances despite their different levels of complexity of each system. On the other hand, the rate achieved by the two manual refactoring changes for Blob type varied considerably \reviewed{based on the human effort of each freelancer, and the complexity of the code that enable (or not) to abstract functionality to new classes. From a practical point of view, developers will not only reduce maintenance costs by using automated approaches, but can ensure a standardized quality gain after the refactoring process.}
 
 \subsection{Improving automated refactoring of Blob classes}
 
To improve the performance of automated refactoring on Blob anti-patterns, it would be necessary to control for code semantics targeting two aspects: (1) the generation of refactoring candidates should be guided by code lexicon; (2) an automated refactoring approach should incorporate a mechanism for naming new code entities based on code lexicon. The first aspect has been already discussed by Ouni~\ea~\cite{Ouni201518}, while the second point remains unexplored. We suggest that by addressing these two aspects we could improve the performance of automated refactoring of Blob classes, making it as good as the human refactorings, or even better.
\subsection{Refactoring changes impact on code comprehension}
Concerning the impact on code comprehension, we could not make an analysis by anti-pattern type as we only have 6 points for each anti-pattern type. However, $E_2$ reveals no significant difference between subject’s efforts, times, and percentages of correct answers on systems refactored by human and machine. We investigated whether the two mitigating variables : expertise and Java's level of confidence (not to be confused with respondent's level of confidence per question in $E_1$) declared by participants impacted our results. We set 4 levels for the years of experience and 5 levels for the degree of Java's confidence, using a Likert scale, from \emph{no confidence} to \emph{highly confident}. \\
Table~\ref{table:mitigatingComprehension} presents some descriptive statistics of the data collected for these two mitigating variables. Since for each mitigating variable we have multiple levels (more than two) corresponding to multiple categories, we used the Kruskal-Wallis Test, which is a non-parametric test for comparing multiple medians, to assess the impact of the mitigating variables on the three performance dimensions (time, effort, and \% of correct answers). We observed that the mitigating variables do not impact our results, as shown by the high $p-values$ in Table~\ref{table:mitigatingComprehension}; \ie{} participants mostly had the same performance on refactorings from the same origin, no matter their level of expertise/Java confidence, which reinforces our assessment that refactoring changes are what did matter.

\begin{table}[t]
\centering
\renewcommand{\arraystretch}{1.0}
\renewcommand{\tabcolsep}{.5mm}
\caption{$p-values$ of the impact of the mitigating variables on the performance of participants for $E_2$}
\label{table:mitigatingComprehension}

\begin{tabular}{|l|l|r|r|r|}
\hline
Variable                        & origin  & \multicolumn{1}{p{0.2\columnwidth}|}{effort: $p-values$} & \multicolumn{1}{p{0.2\columnwidth}|}{time: $p-values$} & \multicolumn{1}{p{0.2\columnwidth}|}{\% of correct answers: $p-values$} \\ \hline
\multirow{2}{*}{Expertise}      & human   & 0.3658                                & 0.9516                              & 0.857                                                \\ \cline{2-5} 
                                & machine & 0.2172                                & 0.6145                              & 0.9109                                               \\ \hline
\multirow{2}{*}{Java confidence} & human   & 0.4522                                & 0.01657                             & 0.346                                                \\ \cline{2-5} 
                                & machine & 0.8893                                & 0.8166                              & 0.1339                                               \\ \hline
\end{tabular}
\end{table}





\section{Threats to Validity}
\label{sec:threats}
There are threats that limit the validity of this study. We discuss these threats and how we alleviate or accept them following common guidelines provided in~\cite{wohlin2012experimentation}.

\subsection{Construct Validity}
Construct validity threats concern the relation between theory and observations. 
\reviewed{This work relies on good developers practices (\ie~extreme programming~\cite{Beck:2004:EPE:1076267}) where developers are advised to perform refactoring to remove anti-patterns, in order to maintain the design quality at an acceptable levels, and hence ease software evolution. However, we cannot claim that removing anti-patterns is the prime reason for developers to perform refactoring, specifically the ones that we studied.  However, relying on the notion of refactoring anti-patterns allowed us to objectively evaluate two methods (manual and automated refactoring), and to control from a large space of possibilities. That is why we had to constraint our study of the refactoring practice to existing well-known refactorings~\cite{fowler1999refactoring,Brown98-AntiPatterns}.}

In $E_2$, we use time and percentage of correct answers to measure the subjects' performance. The measured time was extracted from the video recording of the comprehension tasks sessions, while the percentage of correct answers was evaluated by one author of the paper and one Master's student. We believe that these measurements are objective, even if they can be affected by external factors, such as fatigue. We also use NASA TLX score to measure the subjects' effort. The TLX is by its own nature subjective and, thus, it is possible that our subjects declared effort values that do not perfectly reflect their effort. 

The degree of severity of the anti-patterns is also a threat to construct validity. The anti-patterns instances selected in each system were validated through a voting process for decisions. The first author and a Master's student voted for the anti-patterns, and the second author reviewed the decisions. We based our selections on the definitions and examples provided by Brown and Fowler~\cite{Brown98-AntiPatterns,fowler1999refactoring}. To validate the solutions proposed by the automatic tool and the developers (\ie{} the freelancers), we check that the solution preserves code's behavior based on the unit tests included in the SF110 corpus. We also control for the level of complexity of the refactoring changes proposed by freelancers and our tool by computing normalized change entropy metric, which showed that the changes judged by human evaluators are fair for both treatments~\cf{\ref{table:systems}}.  Yet, it is possible that some of the refactoring changes proposed would have a different effect if applied to other systems in different contexts. 

Construct validity threats could be the result of a mistaken relation between (automated) refactoring and program comprehension. We believe that this threat is mitigated by the fact that this relation seems rational. The results of our analysis suggest that certain anti-patterns' type can be automatically refactored with the same level of quality as the refactorings performed by human developers. 

\subsection{Internal Validity}
We identify 4 threats to the internal validity of our study: learning, selection, instrumentation, and diffusion.
\subsubsection{Learning} Learning threats do not affect our study for the two experiments because we used a between-subject design. A between-subject design uses different groups of subjects, to whom different treatments are assigned. Additionally, we took each anti-pattern instance from different systems. For $E_1$, we balanced the groups (alternating difficulty level, and origin); then we randomized the appearance order of the refactoring changes for each group. For $E_2$, the freelancers had to perform two comprehension tasks. To mitigate the learning effect, the systems were presented in the same order for both treatments (manual and automatic refactoring). For example, consider a comprehension task for systems 47, and 52. We anticipated  that developers will spent more time for the first system (\ie~47), while getting familiar with the instructions, developer environment, etc., than with the second system (52). Hence, the extra time spent as a consequence of the learning process is considered in the same system for both treatments.

\subsubsection{Participant's selection} Participant's selection threats could impact our study due to the natural difference among the participants' skills. In $E_1$, we tried to mitigate this threat by inviting developers through technical Java developers mailing list, (\eg~openJDK project), developers groups on social networks (\eg~LinkedIn, Reddit, and Facebook). In $E_2$, we studied the possible impact of their expertise in Java through two mitigating variables and found no significant 
impact on the obtained results.

\subsubsection{Instrumentation} Instrumentation threats were minimized by using objectives measures like time and percentage of correct answers. We observed some subjectivity in measuring refactoring changes quality, developer's confidence, and developer's experience in $E_1$; and developers' effort measured using the TLX score. For example, 5 years of experience of one developer, could be the equivalent of 3 years for another one. However, this subjectivity is inevitable in self-evaluations.

\reviewed{Another instrumentation threat to our study is the anonymization of new code lexicon.  This only affects Blob, Long-parameter list and Introduce-parameter object anti-pattern types. Automated refactorings that introduce new elements are likely to be distinguished from their manual equivalents. But naming code lexicon is just one part of the semantic context. By anonymizing the names of newly created entities, we wanted to steer the focus of the respondents toward the structure of the code changes. 
Recent works~\cite{hu2018deep} in automated code comment generation have shown promising results on generating human-like comments using deep learning. 
Hence, we believe that there is a reasonable possibility to overcome the current code lexicon limitations of automated refactoring tools in the future, and this study can serve as base for performing further studies when the technology is mature enough.}   

\subsubsection{Diffusion} Diffusion threats do not impact our study because (1) we recruit participants through web platforms and mailing lists, and they do not have physical interactions, and (2) we asked them to not disclose any information about the content of the surveys and the systems.

\subsection {Conclusion Validity}
Conclusion validity threats concern the relation between the treatment and the outcome. We paid attention not to violate the assumptions of the statistical tests that were performed. Indeed, we used non-parametric tests that do not require to make assumptions about the distribution of the data set.

\subsection {Reliability Validity}
Reliability validity threats concern the possibility of replicating this study. We provide all the necessary details to replicate our study in our lab's Web page~\cite{refturingReplication}, including a sample of the questionnaire and the comprehension tasks, and raw data to compute the statistics. The systems analyzed from SF110 are open-source and can be downloaded from the author’s web site. Our automated refactoring tool is also available on-line at \url{https://github.com/moar82/RefGen}.

\subsection{External Validity}
We performed our study on 10 different real-world systems belonging to different domains and with different sizes (see~\Cref{table:systems}). Our experimental design, providing a few classes of each system to each participant, is reasonable because, in real maintenance projects, developers perform their tasks on small parts of whole systems and probably limit themselves as much as possible to avoid getting lost in a large code base. In $E_1$ we summarize the refactoring changes using the diff notation, and provide it along with the original source code; allowing developers to spot changes fast, in a readable and standard way, and in case of doubt, to clone the repository to explore the code and--or apply the changes (by applying the diff file as a patch). To mitigate the impact that a \reviewed{lack of} familiarity to diff notations could have on the responses of our study, we explained the notation to participants prior to \reviewed{participating} in the study and provided them multiple examples in a guidelines document. 
\reviewed{We cannot claim that our results can be generalized to other refactoring tools, and to other subjects. To generalize our results, we should implement other approaches different from RePOR.  However, this was not the main objective of this work.  Rather, we want to empirically evaluate if automatically refactored code can be interleaved with human code on  development and maintenance activities of real world systems.}

Our future work includes replicating this study in other contexts, with other subjects, tools, questions, anti-patterns, and software systems.

\section{Conclusions}
\label{sec:conclusions}
Refactoring tool support is conjectured in the literature to be underused due to lack of awareness, and developers reluctance to incorporate machine-generated-code into their code base. To debunk this myth, and foster awareness of automated refactoring, we performed two experiments to evaluate the perceived quality of automated refactorings and their impact on code comprehension. Our results show that developers could not distinguished between automatically generated refactorings and refactorings created by humans, for 3 out of the 5 anti-pattern types studied. Moreover, in general, developers did not prefer refactorings generated by humans over automatically generated refactorings, the only exception being for removing Blob classes. We found no significant difference between the performance of developers performing comprehension tasks on code refactored by developers or by an automatic tool, to remove Blob, Lazy class, and Speculative generality. 
Hence, we conclude that automated refactoring can be as effective as manual refactoring. However, for complex anti-patterns' types like Blob, we suggest that developer's expertise be included in the refactoring process as much as possible. 
In the future work, we plan to enhance automated refactoring with code semantics by leveraging the code lexicon of systems when determining the best candidates to receive functionalities extracted 
from Blob classes, and for automatically naming the new classes, and--or methods introduced during a refactoring process. By doing this, we could generate more natural refactoring solutions, and close the gap between human and machine generated refactorings. 



\balance
\bibliographystyle{spmpsci}

\end{document}